\documentclass[prl,aps,twocolumn,groupaddress]{revtex4-1}
\usepackage{latexsym}
\usepackage{graphicx}
\usepackage{verbatim}
\usepackage{multirow}

\usepackage{amsmath}
\newcommand{\beq}{\begin{eqnarray}}
\newcommand{\eeq}{\end{eqnarray}}
\usepackage{mathrsfs}
\usepackage{float}
\usepackage[usenames, dvipsnames]{color}
\usepackage{mathtools}
\usepackage{slashed}
\usepackage{physics}	
\usepackage{graphicx}   
\usepackage{epstopdf}
\usepackage{subfigure}  
\usepackage{hyperref}   
\usepackage{bbold}
\usepackage{wasysym}
\usepackage{amsmath}

\begin{document}

\title{Vanishing Hall Conductance in the Phase Glass Bose Metal at Zero Temperature}
\author{Julian May-Mann}
\affiliation{Department of Physics, University of Illinois at Urbana-Champaign, Urbana, Illinois, USA}
\author{Philip W. Phillips}
\affiliation{Department of Physics, University of Illinois at Urbana-Champaign, Urbana, Illinois, USA}

\begin{abstract}
Motivated in part by the numerical simulations\cite{ky,kosterlitz1,kosterlitz2} which reveal that the energy to create a defect in a gauge or phase glass scales as $L^{\theta}$ with $\theta<0$ for 2D, thereby implying a vanishing stiffness, we re-examine the relevance of these kinds of models to the Bose metal in light of the new experiments\cite{kapsym,armitage} which reveal that the Hall conductance is zero in the metallic state that disrupts the transition from the superconductor to the insulator in 2D samples.   Because of the particle-hole symmetry in the phase glass model, we find that bosonic excitations in a phase glass background generate no Hall conductance at the Gaussian level. Furthermore, this result persists to any order in perturbation theory in the interactions. We show that when particle-hole symmetry is broken, the  Hall conductance turns on with the same power law as does the longitudinal conductance.  This prediction can be verified experimentally by applying a ground plane to the 2D samples.
\end{abstract}

\maketitle

\medmuskip=0mu
\thinmuskip=0mu
\thickmuskip=0mu

Because of the canonical relationship between phase and particle number, bosons are traditionally thought to either condense in an eigenstate of phase (superconducting) or insulate as dictated by particle number eigenstates.  Indeed, the initial experiments\cite{hebard2,hebard1,paalanen628} seemed to conform to the predictions\cite{mpaf} of the phase-only XY model that only at the critical point do bosons exhibit the  quantum of resistance of $h/4e^2$.  However, subsequent experiments\cite{G1989,yazdani,valles628,vallesmetal} indicated that there is nothing special about the value of the resistance at the critical point, thereby calling into question the relevance or accuracy of the prediction of the phase-only model that only bosons on the brink of localization conduct with the quantum of resistance for charge $2e$ carriers.   More importantly, since 1989\cite{ephron,yoon2006,park2017,iwasa,mason,masonthesis,vallesmetal,vallesgauge,liu2009,G2005,yazdani3,armitage2013,Tsen2015,thicknessbehnia}, a state with apparent finite $T\rightarrow 0$ resistivity appeared immediately upon the destruction of superconductivity.  Although questions of thermometry were raised regarding the initial\cite{G1989} observation, the leveling of the resistance persisted in the magnetic-field tuned transition in MoGe\cite{ephron,mason,yazdani3}, Ta\cite{yoon2006,yoonTa}, InO$_x$\cite{armitage2013,kapsym}, and NbSe$_2$\cite{liu2009,Tsen2015}.  The key contribution of the magnetic-field tuned data was to clarify that the intervening state occurred well below $H_{c2}$.  Consequently, if these observations constitute a true metallic state at $T=0$, the charge carriers must be 2e bosons that lack phase coherence.  As a result, the insulator above $H_{c2}$ is mediated by the breaking of the Cooper pairs.  

 The newer observations of the Bose metal in cleaner samples with either gate\cite{iwasa} or magnetic-field tuning\cite{Tsen2015} tell us three things.  First, in the field-effect transistors\cite{iwasa} composed of ion-gated ZrNCl crystals, the superconducting state which obtains for gate voltages exceeding 4V is destroyed\cite{iwasa} for perpendicular magnetic fields as low as $0.05T$.  The authors\cite{iwasa} attribute this behaviour to weak pinning of vortices and hence reach the conclusion that throughout most of the vortex state, be it a liquid or a glass, a metallic state obtains.  Second, in the NbSe$_2$ samples, essentially crystalline materials,  the resistance turns on\cite{Tsen2015} continuously as $\rho\approx (g-g_c)^\alpha$, where $g_c$ is the critical value of the tuning parameter for  the onset of the metallic state.  Similar results have also been observed in MoGe\cite{masonthesis}.  Third, in InO$_x$ and TaN$_x$, the Hall conductance is observed\cite{kapsym} to vanish throughout the Bose metallic state, thereby indicating that particle-hole symmetry is an intrinsic feature of this  state.  In strong support of this last claim are the recent experiments demonstrating that the cyclotron resonance vanishes in the Bose metallic state\cite{armitage}.

While there have been numerous proposals for a Bose metal\cite{p1,p2,p3,p4,p5,dp2},  a state with a finite resistance at $T=0$, the new experiments highly constrain possible theoretical descriptions.  In light of the new experimental findings, we re-examine the phase glass model we proposed several years ago\cite{dp1,dp2,jw1} which we demonstrated, using the Kubo formula in the collision-dominated (or hydrodynamic) regime, to have a finite $T\rightarrow 0$ resistivity that turns on as $\rho\approx (g-g_c)^\alpha$, as highlighted in the experiments on NbSe$_2$\cite{Tsen2015}.  While questions\cite{stiffness} regarding the phase stiffness of the phase glass have been raised, numerical simulations all indicate\cite{ky,kosterlitz1,kosterlitz2} that the energy to create a defect in a 2D phase or gauge glass scales as $L^{\theta}$, where $\theta=-0.39$.  Hence, the stiffness is non-existent.  In 3D\cite{ky,kosterlitz1,kosterlitz2}, $\theta>0$ and a stiffness obtains.  As such glass states are candidates to explain the vortex glass\cite{fisher,vortexglass},  that $\theta<0$ is consistent with the experimental finding\cite{iwasa} in ion-gated ZrNCl, an extreme 2D system, that the resultant vortex state is indeed metallic and not a true superconductor.   

In this paper, we show that the Hall conductance in the phase glass model vanishes as observed experimentally as a result of an inherent particle-hole symmetry. 

As shown previously\cite{spivakkiv}, any amount of dirt in a 2D superconductor induces $\pm J$ disorder, $J$ the Josephson coupling.  Consequently, a disordered superconductor is closer to a disordered XY model  rather than a dirty superfluid.  Justifiably, the  starting point for analyzing the experiments is the disordered XY model.  Since we wish to calculate the Hall conductance from the Kubo formula, we consider 
\begin{equation}
H = -E_C \sum_i (\frac{\partial}{\partial \theta_i})^2 - \sum_{\langle i,j \rangle} J_{ij} \cos(\theta_i - \theta_j - A_{ij}),
\label{eq:Ham}
\end{equation}
 the phase Glass in a perpendicular magnetic field, where $A_{i j} =  e^*/\hbar\int_i^j \textbf{A} d\textbf{l}$, ($e^* = 2e$), $E_C$ is the constant onsite and $J_{ij}$ is the strength of the Josephson couplings which are randomly distributed according to $P(J_{ij}) = 1 / \sqrt{2 \pi J^2} \exp((J_{ij}-J_0)^2 / 2J^2 )$, with non-zero mean $J_0$. In terms of the phase on each island, we introduce  the vector $S_i = (\cos(\theta_i),\sin(\theta_i))$.    This will allow us to 
 to recast the interaction term
in the random Josephson Hamiltonian as a spin problem
with random magnetic interactions,
 $\sum_{\langle i,j\rangle}J_{ij}{\bf S}_i \cdot {\bf S}_j$.
Let $\langle ...\rangle$ and $[...]$ represent
averages over the thermal degrees of freedom
and over the disorder, respectively.  
In the superconductor not only $\langle S_{i\nu}\rangle$
but also $[\langle S_{i\nu}\rangle]$
acquire a non-zero value. In the phase (or spin) glass, however, 
$\langle S_{i\nu}\rangle\ne 0$ but $[\langle S_{i\nu}\rangle ]=0$.  As we have shown previously\cite{dpprb,rsy}, the Landau theory for this problem is obtained by using
replicas to average over the disorder and the identity 
$\ln [Z]=\lim_{n\rightarrow 0}([ Z^n]-1)/n$ to obtain the zero 
replica limit.  The quartic and quadratic spin-spin
interaction terms that arise from the disorder average can be decoupled
by introducing the auxiliary real fields,
\beq
 Q_{\mu\nu}^{ab}(\vec k,\vec k',\tau,\tau')=\langle
S_\mu^a(\vec k,\tau)S_\nu^b(\vec k',\tau')\rangle
\eeq
and $\Psi^a_\mu(\vec k,\tau)=\langle S^a_\mu(\vec k,\tau)\rangle$, 
respectively. Here the superscripts indicate the replica indices and the subscripts the components of the spin. To simplify our notation, we will introduce the one component complex field $\psi^a = (\Psi^a_1,\Psi^a_2)$.
 
Taking into account the effects of the magnetic field $\textbf{B} = B \hat{z}$, we will use Landau Gauge $\textbf{A} = (0,Bx,0)$, and rewrite $\psi$ as a sum over different landau levels
\begin{equation}
\psi^a ( l,x,y,\omega, p_y) = C^a_{l,p_y}(\omega) \phi_l(x-\frac{\hbar p_y}{e^* B})e^{i p_y y},
\label{eq:land}
\end{equation}
where $\phi_l$ is the normalized eigenstate of the harmonic oscillator. 
The relevant part of the free energy consists of the purely bosonic degrees of  freedom and their coupling to the phase glass sector which is controlled by the Edwards-Anderson order parameter.  This free-energy has been derived previously\cite{dp2,dpprb}.  To tailor the expressions to a calculation of the Hall conductance, we expand the $\psi$ degrees of freedom in terms of the Landau levels.  The resulting free energy per replica is then 
\beq
\mathcal{F}_{\psi}[C,Q] &=& \sum_{\substack{a,l,p_y,\omega_n}} (m_H^2(l+\frac{1}{2})+\omega_n^2 + m^2)|C^a_{l,p_y}(\omega_n)|^2 \nonumber\\
 &- &\frac{1}{\kappa t}\sum_{\substack{a,b,l\\p_y,\omega_n,\omega'_n}} C^a_{l,p_y}(\omega_{n})C^{b*}_{l,p_y}(\omega'_{n}) Q^{ab}(l,p_y,\omega_{n},\omega'_{n})\nonumber\\
&+&\frac{U}{2} \sum_{a,l_i,\omega_{ni},p_{yi}}
|\psi^a ( l_i,x,y,\omega_{ni},p_{yi} )|^4
\eeq
where $C^a_{l,m}$ describes bosonic excitations with charge $2e$, $\kappa$, $t$, and $U$ are standard Landau theory parameters\cite{dp2,rsy}, $m^2$ is an inverse correlation length, $\omega_n$ are the Matsubara frequencies, and $m_H^2 = \frac{e^*}{c \hbar}B$. We have left the interaction in terms of $\psi$ for the simplicity, with $i = (1,2,3,4)$. There is also a contribution to the free energy from terms only proportional to $Q$. In our analysis we will only be treating the $\psi$ field dynamically, and so these terms can be ignored. 
As shown previously\cite{dp1,rsy} the spin-glass order parameter is of the form,
\beq
Q^{ab} (l,p_y,\omega_1,\omega_2) &=& \beta (2\pi)^2 \delta_{l,0}\delta_{p_y,0} [-\eta|\omega_1|\delta_{\omega_1+\omega_2,0}\delta^{ab} 
\nonumber\\ &+& \beta \delta_{\omega_1,0}\delta_{\omega_2,0} q^{ab}]
\label{eq:Q2}
\eeq
where $\eta = 1/\kappa^2\tau$ and $q^{ab}$ is a symmetric ($q^{ab} = q$ for all $a,b$) ultrametric matrix. Due to the factor of $|\omega_n|$ the dynamic critical exponent of this system is $z = 2$ and as a result particle-hole symmetry is a natural consequence.

Substituting Eq. (\ref{eq:Q2}) into the free energy, we obtain
\beq
\mathcal{F}_{\psi}[C] &=& \sum_{a,l,m,\omega_n} (m_H^2(l+\frac{1}{2})+\omega_n^2 + \eta|\omega_n|+ m^2)|C^a_{l,p_y}(\omega_n)|^2 \nonumber\\ 
&-& \beta q^{ab}\sum_{a,b,l,p_y,\omega_n} C^a_{l,p_y}(\omega_{n})C^{b*}_{l,p_y}(\omega_{n})\nonumber\\
&+& \frac{U}{2} \sum_{a,l,p_y}
|\psi^a ( l,x,y,p_y )|^4,
\label{eq:Free2}
\eeq
where we have shifted $q \rightarrow q\kappa t$.  
The propagator for the Gaussian part of the theory is given by 
\beq
G^{ab}(l,p_y,\omega_n) &=& G_{0}(l,\omega_n)\delta^{ab} + \beta G^2_{0}(l,\omega_n)q^{ab}\nonumber\\
G_{0}(l,\omega_n) &=& (m_H^2(l+\frac{1}{2})+\omega_n^2 + \eta|\omega_n|+ m^2)^{-1}.
\label{eq:Prop1}
\eeq
As is well known, $G_0$, the propagator in the presence of  Ohmic dissipation\cite{chakrav,dp2000} is insufficient to describe the metallic state.  Such physics originates from the characteristic double-trace deformation the spin-glass term induces in the full Gaussian propagator,  $G^{ab}$.    Note that $G(l,\omega_n)$ is symmetric under $\omega_n \rightarrow -\omega_n$. This will be referred to as particle-hole symmetry from here on. 

To find the Hall Conductance for this system, we will use the Kubo Formula 
\begin{equation}
\begin{split}
\sigma_H(i \omega_\nu) = \sigma_{xy}(i \omega_\nu) = \frac{\hbar}{\omega_\nu}\int d^2(x-x') \int d(\tau-\tau') \\
\times \frac{\partial^2 [Z^n]}{\partial A_x(x,\tau) \partial A_y(x',\tau')} e^{i \omega_\nu (\tau-\tau')}.
\end{split}
\label{eq:Kubo0}
\end{equation}
For our system, this simplifies to
\beq
\sigma_H(i \omega_\nu) &=& \frac{i(e^{*}m_H)^2}{2 \omega_\nu \hbar \beta}  \sum_{\substack{a,b,l,l',p_y,p_y',\\p_y'',\omega_n,\omega_n',\omega_n''}}  \int d \tau e^{i\omega_\nu \tau} \sqrt{(l+1)(l'+1)}\nonumber\\ 
&\times& \langle(C^a _{l,p_y}(\omega_n) C^{a*} _{l+1,p_y}(\omega_n)+C^a _{l+1,p_y}(\omega_n) C^{a*} _{l,p_y}(\omega_n))\nonumber\\ 
&\times& (C^b _{l',p_y'}(\omega'_n) C^{b*} _{l'+1,p_y''}(\omega''_n)-C^b _{l'+1,p_y'}(\omega'_n) C^{b*} _{l',p_y''}(\omega''_n))\rangle .\nonumber\\ 
\label{eq:Kubo1}
\eeq
At the Gaussian level, $p_y$ has no effect, and so for the following calculations we will suppress it.
Using Eq. (\ref{eq:Prop1}) we then have
\beq
\sigma_{H}(i \omega_\nu) &=& \frac{i(e^{*}m_H)^2}{2 \omega_\nu \hbar \beta}  \sum_{a,b,l\omega_n} (l+1)\nonumber\\&\times&[G^{ab}(l,\omega_n)G^{ab}(l+1,\omega_n + \omega_\nu) \nonumber\\ &-& G^{ab}(l+1,\omega_n)G^{ab}(l,\omega_n + \omega_\nu)].
\label{eq:Kubo2}
\eeq
It has already been shown that in an array of Josephson Junctions without random couplings, the Hall Conductance vanishes as $T \rightarrow 0$ \cite{van1993response}. As such we will only be considering terms in Eq. ~(\ref{eq:Kubo2}) which are proportional $q$.  Since $q^{ab}q^{ab}$ vanish in the $n \rightarrow 0$ limit, we will only consider terms involving $q^{ab}\delta^{ab} = q$. After using the delta functions to sum over $\omega_n$ we then have for the Hall Conductance 
\beq
\sigma_H&=& 
\frac{i(e^{*}m_H)^2}{2 \omega_\nu \hbar }  \sum_{a,b,l}  (l+1)q[ G_0(l,-\omega_\nu) G^2_0(l+1,0) \nonumber\\ &+&   G_0(l+1,\omega_\nu) G^2_0(l,0) - 
G_0(l+1,-\omega_\nu) G^2_0(l+1,0) \nonumber\\ &-&  G_0(l,\omega_\nu) G^2_0(l+1,0)].\nonumber\\
\label{eq:Kubo3}
\eeq
Due to the particle-hole symmetry of the propagator, the Hall conductance then vanishes independent of $\omega_\nu$ and $\lim_{\omega_\nu \rightarrow 0} \sigma_H({i\omega_\nu})= 0$. After taking the $T \rightarrow 0$ limit, we conclude that at the Gaussian level, the Hall conductance at $T = 0$ vanishes. 

We now look at the role of interactions. To do this, we will consider the exact propagator, the exact 4-point vertex, and show that even with interactions, the Hall conductance still vanishes as $T \rightarrow 0$. In doing this, we are assuming that all effects can be re-summed into a new propagator and a new vertex.  Taking into account the quartic interaction term in the free energy, we rewrite the exact propagator in the $n \rightarrow 0$ limit in the form 
\begin{equation}
\mathcal{G}^{ab}(l,\omega_n,p_y) = \widetilde{G}(l,\omega_n,p_y) \delta^{ab} +\beta q^{ab} g(l,p_y) \delta_{\omega_n,0},
\label{eq:exctprop}
\end{equation}
where we have split the propagator into a diagonal component involving $\widetilde{G}$ and and off-diagonal component involving $g$\cite{dp1}. Since all diagrams have particle hole symmetry, we can conclude that $\widetilde{G}$ must also have particle hole symmetry. Since $\beta$ only couples to the diagonal component in the original free energy, we conclude that the diagonal components of the Eq. ~(\ref{eq:exctprop}) must be independent of $\beta$. 

\begin{figure}[h]
\includegraphics[width=\linewidth]{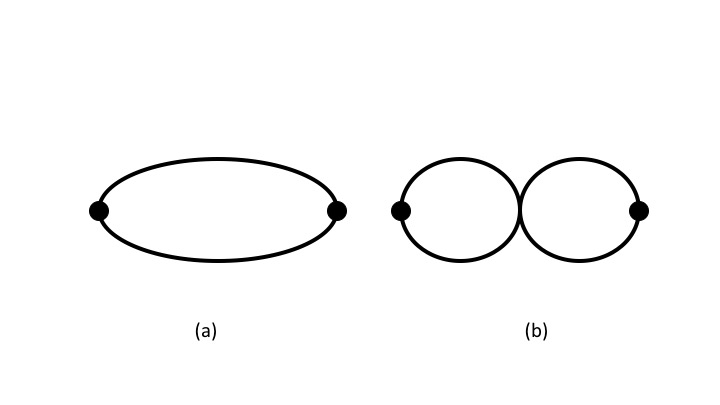}
\caption{(a) The non-vertex diagram and (b) vertex diagram that contribute to the Hall Conductance. Here the propagators are the exact propagator and the vertex is the exact vertex. }
\label{fig:diaexct}
\end{figure}

The exact 4-point propagator is given by
\beq
\Gamma(l_i,\omega_j,p_{yk}) &=& \frac{U}{\beta} \sum_{l_1,\omega_j,p_{yk}}\delta_{\Sigma \omega_i,0} \delta_{\Sigma p_{yk},0} f(l_i,\omega_j, p_k) \nonumber\\&\times& C_{l_1,p_{y1}} (\omega_1) C^*_{l_2,p_{y2}}(\omega_2) C_{l_3,p_{y3}}(\omega_3)C^*_{l_4,p_{y4}}(\omega_4),\nonumber \\
\label{eq:exctvert}
\eeq
where the exact form of $f(n_l,\omega_j,p_{yk})$ is unknown, but since any diagram can be rotated or switched, we expect $f$ to be independent of the order of its parameters\cite{dp1}. Due to the particle-hole symmetry of all diagrams, we can also conclude that $f(n_i,\omega_j) = f(n_i,-\omega_j)$. 
If we use the exact propagator and the exact vertex, there are two diagrams that contribute to $\sigma_H(i\omega_\nu)$; see Fig ~(\ref{fig:diaexct}).

The diagram in Fig.  ~(\ref{fig:diaexct}) (a) has a contribution of 
\beq
\sigma_{H1} &=& \frac{i(e^{*}m_H)^2}{2 \omega_\nu \hbar \beta}  \sum_{a,b,l\omega_n,p_y} [\mathcal{G}^{ab}A_1(l)(l,\omega_n,p_y)\nonumber\\
&\times& \mathcal{G}^{ab}(l+1,\omega_n + \omega_\nu,p_y)] - [\mathcal{G}^{ab}(l+1,\omega_n,p_y)\nonumber\\ &\times&\mathcal{G}^{ab}(l,\omega_n + \omega_\nu,p_y)]\nonumber\\
\label{eq:exctdia1}
\eeq
where $A_1(l)$ is a dimensionless function of $l$. Using Eq. ~(\ref{eq:exctprop}), we can expand Eq.  ~(\ref{eq:exctdia1}) into terms proportional to $\delta^{ab}\delta^{ab}$, terms proportional to $\beta q^{ab}\delta^{ab}$ and terms proportional to $\beta^2 q^{ab}q^{ab}$. The terms proportional to $\delta^{ab}\delta^{ab}$ vanish in the $T \rightarrow 0$ limit and the $\beta^2 q^{ab}q^{ab}$ terms vanish in the $n \rightarrow 0$ limit (which is taken before the $T \rightarrow 0$ limit). So the only terms remaining are proportional to $\beta q^{ab}\delta^{ab} = \beta q$. Evaluating these terms, we find that, due to the particle hole symmetry of $\widetilde{G}$ this diagram does not contribute to the Hall conductance. 

Similarly the diagram in ~(\ref{fig:diaexct}) (b) diagram yields
\beq
\sigma_{H2} &=& \frac{U(e^{*}m_H)^2}{2 \omega_\nu \hbar \beta^2} \sum_{\substack{a,b,l,l'\\ \omega_{n},\omega_n,p_y,p_y'}} A_2(l,l')f(l,l',\omega_n,\omega_n',p_y,p_y') \nonumber\\ &\times& \mathcal{G}^{ab}(l,\omega_n,p_y)\mathcal{G}^{ab}(l+1,\omega_n + \omega_\nu,p_y)\nonumber\\&\times&\mathcal{G}^{bc} (l'+1,\omega'_n,p_y')\mathcal{G}^{bc}(l',\omega'_n + \omega_\nu,p_y') \nonumber\\&+& \mathcal{G}^{ab}(l+1,\omega_n,p_y)\mathcal{G}^{ab}(l,\omega_n + \omega_\nu,p_y)\nonumber\\&\times& \mathcal{G}^{bc} (l'+1,\omega'_n,p_y')\mathcal{G}^{bc}(l',\omega'_n + \omega_\nu,p_y'),\nonumber\\
&-&(l' \leftrightarrow l'+1),
\label{eq:exctdia2}
\eeq
where $A_2(l,l')$ is dimensionless and is symmetric in $l$, and $l'$. For this diagram, let us first fix the values of $l$ and $l'$. If we then expand the propagators as we did with the propagators in the first diagram and invoke particle-hole symmetry of $f$ and $\mathcal{G}$, we find that the contribution from each $l$ and $l'$ vanishes independently of $\omega_\nu$. Thereby, the second diagram also does not contribute to the Hall conductance. From this we can conclude that the Hall conductance will remain 0 at all levels in perturbation theory.  Consequently, the phase Glass model of the Bose metal is consistent with the vanishing of the Hall conductance even in the presence of interactions. In Apendix A we explicitly carry out the calculations to linear order in U and show that these contributions vanish. 

We now consider the effects of breaking the particle-hole symmetry of this system. This can be done by including a term $i\lambda \psi^* \partial_\tau \psi$ in the free energy. This changes the propagator to 
\begin{equation}
G_{0}(l,\omega_n,p_y) = (m_H^2(l+\frac{1}{2})+\omega_n^2 + \eta|\omega_n|+ i\lambda\omega_n + m^2)^{-1}.
\label{eq:propbr}
\end{equation}
This term breaks particle-hole symmetry. Without the effects of dissipation ($\eta = 0$), the number of particles at finite temperature is given by $N(\omega^{\pm}) = [\exp(\beta \omega^{\pm}) - 1]^{-1}$, where $\omega^{\pm} = \mp \lambda + \sqrt{\lambda^2+m^2}$\cite{van1993response}. Particle-hole symmetry is thereby restored at $\lambda = 0$. 

We will first look at the Hall conductance of this system at the Gaussian level (again suppressing $p_y$). 
Using the Kubo Formula, Eq.  ~(\ref{eq:Kubo2}), we find that 
\beq
\sigma_H({i\omega_\nu}) &=& \frac{\lambda(e^{*}m_H^2)^2}{ \hbar }  \sum_{a,b,l}  q (l+1)[G_0(l,-\omega_\nu) G_0(l,\omega_\nu) \nonumber \\&\times& G^2_0(l+1,0) + G_0(l+1,-\omega_\nu) G_0(l+1,\omega_\nu) G^2_0(l,0)]\nonumber\\ 
\label{eq:Kubo1br}
\eeq
Taking the limits $\omega_\nu \rightarrow 0$ and $T \rightarrow 0$, and then evaluating the sum, we find
\begin{equation}
\begin{split}
\sigma_H({i\omega_\nu}) = \frac{\lambda q (e^{*}m_H^2)^2}{ \hbar m^4 } (\frac{2}{x} - \frac{\Psi(1,\frac{x+2}{2x})}{x^3} ),
\end{split}
\label{eq:result1br}
\end{equation}
where $x = \frac{m^2_H}{m^2}$, and $\Psi(1,x)$ is the first digamma function. 
In the low magnetic field regime ($x \ll 1$), the Hall conductance is approximately
\begin{equation}
\begin{split}
\sigma_H({i\omega_\nu}) = \frac{\lambda q 4 e^{2*}}{ 3 \hbar m^4 } (1 + \frac{m_H^2}{m^2} ).
\end{split}
\label{eq:result2br}
\end{equation}
In the high magnetic field regime ($x \gg 1$), the Hall conductance is 
\begin{equation}
\begin{split}
\sigma_H({i\omega_\nu}) = \frac{\lambda q  e^{*2}}{ \hbar m_H^4 } (2 + \frac{\pi^2 m^2}{m_H^2} ).
\end{split}
\label{eq:result3br}
\end{equation}

So in the case of broken particle-hole symmetry, there is a non-vanishing Hall conductance for all ranges of the magnetic field. The Hall conductance also scales algebraically throughout this range. This is contrary to the results from a nonrandom array of Josephson junctions, where it was shown that the Hall conductance vanishes when $T \rightarrow 0$ even in the case of a broken particle-hole\cite{van1993response} term. 

In the presence of a broken particle hole symmetry, the longitudinal conductance of the of this system is given by 
\begin{equation}
\begin{split}
\sigma_{xx}({i\omega_\nu}) = \frac{\eta q (e^{*}m_H)^2}{ \hbar m^4 } (\frac{2}{x} - \frac{\Psi(1,\frac{x+2}{2x})}{x^3} ),
\end{split}
\label{eq:result1}
\end{equation}
which is unchanged from the particle hole symmetric case \cite{dp1}.
Thus if both $\lambda \neq 0$ and $\eta \neq 0$, we see that the longitudinal and Hall conductances have the same algebraic scaling, a falsifiable prediction of this theory.

We will now look at corrections to the Hall conductance arising from quartic interactions at linear order in $U$. 
\begin{figure}[h]
\includegraphics[width=\linewidth]{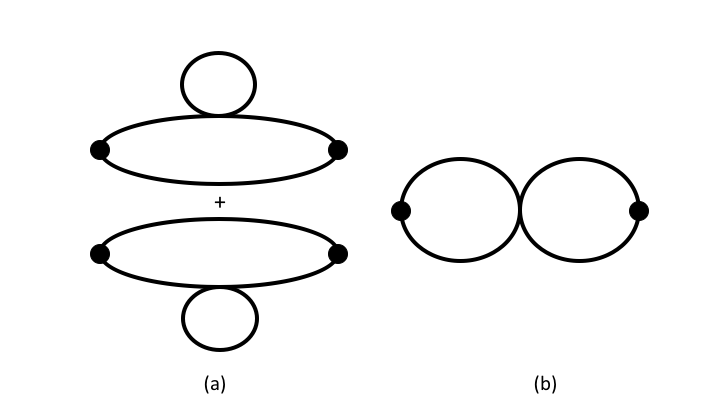}
\caption{The two diagrams that contribute to the Hall Conductance in the presence of a broken particle-hole symmetry: (a) The loop correction that which will be expressed as a rescaling of the mass and (b) the vertex correction. 
The propagator shown here is at the Gaussian level and the vertex is that which appears in the free energy. }
\label{fig:diabr}
\end{figure}
These contributions come from the following diagrams in Fig. ~(\ref{fig:diabr}). The effects from the Fig. ~(\ref{fig:diabr}) (a) can be expressed as a redefinition of the $m$ term giving in the $\omega_\nu \rightarrow 0$ and $T \rightarrow 0$ limit \cite{jw1} limit,
\begin{equation}
\widetilde{m}^2 = m^2 +  \frac{U q m_H^2}{4\pi} \sum_l (G^{(0)}_{l,0})^2 = \frac{U q }{4\pi m_H^2 } \Psi(1,x+1/2).
\label{eq:massRG}
\end{equation}
For the high magnetic field regime, the correction to the mass in Eq.~(\ref{eq:massRG}) is approximately $\frac{\pi U q}{8 m_H^2}(1 + \frac{\Psi(2,1/2)m^2}{2\pi m_H^2})$, and in the low magnetic field regime it is  $\frac{U q}{4 \pi m^2}$.

To evaluate Fig. ~(\ref{fig:diabr}) (b), we will use Eq. \ref{eq:land} to write the interaction term as 
\beq
\Gamma &=& \frac{U}{2} \sum_{l_i,\omega_j,p_{yk}} C^a_{l_1,p_{y1}}(\omega_1)C^{*a}_{l_2,p_{y2}}(\omega_2)C^a_{l_3,p_{y3}}(\omega_3)C^{*a}_{l_4,p_{y4}}(\omega_4)\nonumber\\&\times& \phi_{l_1}(x-\frac{\hbar p_{y1}}{e^* B})\phi_{l_2}(x-\frac{\hbar p_{y2}}{e^* B})\phi_{l_3}(x-\frac{\hbar p_{y3}}{e^* B})\phi_{l_4}(x-\frac{\hbar p_{y4}}{e^* B})\nonumber\\&\times&\delta_{\Sigma \omega_j,0}\delta_{\Sigma p_{yk},0}.\nonumber\\
\label{eq:vertexRG}
\eeq
Inserting Eq. (\ref{eq:vertexRG}) in Eq. (\ref{eq:Kubo1}) and evaluating the sums and integrating we find that the contribution is zero due to the orthogonality of $\psi_l$ and $\psi_{l+1}$. This calculation will be explicitly done in Appendix A. As a result, to linear order in $U$ the only correction to the Hall conductance comes from the rescaling of the mass term. 

In conclusion we have shown that the vanishing of the Hall conductance found in experiments is consistent with the phase glass model. Even if interactions are considered, the Hall conductance remains zero. This is a consequence of the fact the system obeys a particle-hole symmetry. However, if the particle-hole symmetry is broken explicitly we see that there is a non-vanishing Hall conductance that persists even at zero temperature. This finite Hall conductance at zero temperature is a result of both breaking the particle hole symmetry and the glassy nature of the system. Furthermore, the Hall conductance scales the same way as the longitudinal conductance of this system and the phase glass with particle hole symmetry.  This falsifiable prediction can be confirmed by ground-plane experiments and should offer a new window into the true nature of the ground state of the Bose metal.

\textit{Acknowledgments:} We thank Steve Kivelson for pointing out Ref. 1 and the NSF DMR-1461952 for partial funding of this project.

%

\end{document}